# Phase Separation and Charge Transfer in a K-doped $C_{60}$ Monolayer on Ag(001)

M. Grobis,* R. Yamachika, A. Wachowiak, Xinghua Lu, and M. F. Crommie

University of California at Berkeley, Berkeley, California 94720-7300, USA and

Materials Sciences Division, Lawrence Berkeley Laboratory, Berkeley, California 94720-7300, USA

We have performed a scanning tunneling microscopy and spectroscopy study of potassium-doped $C_{60}$ monolayers ($K_xC_{60}$) on Ag(001) in the regime of $x \approx 1$. Low temperature annealing (640 K) leads to the formation of two well-ordered $K_xC_{60}$ phases that exhibit differing levels of electron charge transfer. Further annealing (710 K) distills out the higher electron-doped phase from the lower electron-doped phase, leaving behind a third $C_{60}$ phase completely devoid of K. Spectroscopic measurements indicate that the electron-doping level of the higher electron-doped $KC_{60}$ phase is anomalously large.

Molecular $C_{60}$ exhibits a unique combination of electronic, vibrational, and chemical properties that allow it to function as a key component in materials such as organic solar cells, single electron transistors, and molecular superconductors [1]. $C_{60}$'s versatility stems in part from the ability to tune its electronic structure by doping with electron donors such as alkali atoms [2]. In monolayer systems the substrate-$C_{60}$ interaction provides an additional channel for charge transfer and also serves as an important driving force in determining film morphology. Photoemission experiments performed on $K_xC_{60}$ monolayers on various metallic substrates have revealed a diverse range of electronic structure properties in these systems [3-6]. The interpretation of such spatially averaged $K_xC_{60}$ measurements, however, is complicated by a lack of knowledge about phase separation dynamics [4, 7] in these monolayer systems, as well as discrepancies between observed substrate-$C_{60}$ charge transfer using different techniques [6, 8-10]. The principles guiding the morphology and electronic structure of $K_xC_{60}$ monolayer systems are complex and not completely understood. Local probe measurements are a useful technique for addressing these issues, due to their ability to resolve geometric and electronic structure at molecular length scales [4, 9-12].

Here we present a scanning tunneling microscopy (STM) and spectroscopy (STS) study of the morphology and electronic structure of a K-doped $C_{60}$ monolayer on Ag(001) in the low-doping regime where the average potassium content is approximately one K per $C_{60}$ ($K_xC_{60}$, $x \approx 1$). We observe pronounced temperature-dependent surface morphology changes and phase separation in this monolayer system. While lightly annealed films exhibit a high degree of geometric and electronic disorder, annealing to higher temperatures leads to the formation of two distinct and well-ordered $C_{60}$ phases, here referred to as the $C_{60}^\alpha$ phase and $C_{60}^\beta$ phase. Further annealing causes the $C_{60}^\alpha$ phase to distill out of the $C_{60}^\beta$ phase, resulting in a monolayer composed of well-separated $C_{60}^\alpha$ phases and K-free $C_{60}$ regions, accompanied by a complete disappearance of the $C_{60}^\beta$ phase. The α and β monolayer phases have a similar K concentration, which lies in the range of 0.5 to 1.5 K atoms per $C_{60}$. Local spectroscopic measurements, however, imply a larger than expected $C_{60}$ charge state following K addition. Our data suggests that in addition to electron doping, K affects other aspects of the Ag-$C_{60}$ charge transfer and local electron density distribution. These results point toward an explanation of observed discrepancies between scanned probe and photoemission measurements of molecular monolayer charge states [6, 8-10].

Our experiments were conducted using a homebuilt ultra-high vacuum (UHV) STM with a PtIr tip operated at 7 K. The single-crystal Ag(001) substrate was cleaned via sputter/anneal cycles in UHV and cooled to ~80 K before deposition of potassium using a K getter manufactured by SAES Getters. The K deposition was calibrated by counting the deposited K atoms via STM, prior to depositing $C_{60}$ onto the sample at ~80 K via a $C_{60}$ Knudsen cell. For these studies, the average monolayer stoichiometry ranged from $K_{0.5}C_{60}$ to $K_{1.5}C_{60}$. The K-doped $C_{60}$ sample was progressively annealed to higher temperatures between STM measurements. All anneals consisted of a five minute ramp from room temperature to the stated temperature (± 15 K) whereupon heating was stopped after 1 to 5 minutes [13]. *dI/dV* spectra and images were measured through lock-in detection of the ac tunneling current driven by a 450 Hz, 1-10 mV (rms) signal added to the junction bias under open-loop conditions (bias voltage here is defined as the sample potential referenced to the tip).

The morphology of the K-doped $C_{60}$ film after annealing to 415 K is shown in Fig. 1(a). $C_{60}$ molecules congregate into patches of brighter and dimmer molecules and the film does not show the long range order observed previously in undoped $C_{60}$ monolayers on Ag(001) [9, 12]. $C_{60}$ packing density, height, and orientation all vary between patches. The $C_{60}$ molecules show some preference to locally orient their 5-6 or 6-6 bonds along the Ag(001) [-110] and [110] crystallographic axes, as seen previously for undoped monolayers on Ag(001) [9]. Representative STS spectra of the $C_{60}$ molecules in the 415 K annealed film are shown in Fig. 1(b). Each of these spectra show three distinct resonances associated with the $C_{60}$ highest occupied molecular orbital (HOMO), lowest unoccupied molecular orbital (LUMO), and LUMO+1. Bright $C_{60}$ molecules have LUMO and LUMO+1 resonances at slightly higher energies than dim molecules.

Fig. 2(a) shows the morphology of the same K-doped $C_{60}$ film annealed to a higher temperature of 640 K. Here the film shows a dramatic separation of the $C_{60}$ monolayer into two highly ordered phases, which we denote as the $C_{60}^{\alpha}$ and $C_{60}^{\beta}$ phases. The two phases differ in three ways: (1) topographic height, (2) lattice structure, and (3) electronic structure. The average height of the $C_{60}^{\beta}$ molecules appears ~1.5 Å higher than the $C_{60}^{\alpha}$ phase when measured at $V = 2.0$ V, $I = 50$ pA. Nearly all of the $C_{60}^{\beta}$ phase molecules are arranged in a hexagonal lattice with the molecules oriented so that a hexagon ring faces up. The $C_{60}^{\alpha}$ phase molecules are arranged in a c(6x4) lattice with two orientations: 6-6 bond up and apex atom up. The electronic structure of the $C_{60}^{\alpha}$ and $C_{60}^{\beta}$ phases is shown in the representative STS dI/dV spectra of Fig. 2(b). The LUMO and LUMO+1 resonances of $C_{60}^{\alpha}$ are shifted to lower energy with respect to the Fermi energy ($E_F$) compared to $C_{60}^{\beta}$. The shoulder of the HOMO resonance appears shifted to higher energy for the $C_{60}^{\alpha}$ spectrum (possibly due to the voltage dependence of the tunneling barrier height [14]).

The K-doped $C_{60}$ film undergoes further morphological changes when the annealing temperature is raised to 710 K, as seen in Fig. 3(a). The $C_{60}^{\beta}$ phase completely disappears and two well-separated phases dominate the surface. The first of these phases (bottom right panel of Fig. 3(a)) is the $C_{60}^{\alpha}$ phase seen previously, but which now constitutes a larger fraction of the $C_{60}$ film. The second phase (top right panel of Fig. 3(a)) is undoped $C_{60}$ (i.e., no K). The undoped $C_{60}$ phase contains an arrangement of bright and dim molecules in a hexagonal lattice and is identical in appearance to undoped $C_{60}$ monolayers observed previously on Ag(001) [9, 15, 16]. STS spectra of the two phases are shown in Fig. 3(b). Comparison of the spectra in Fig. 3(b) with Fig. 2(b) and Ref. [9] confirms the identification of the two phases. These results indicate that high-temperature annealing of the K-doped $C_{60}$ monolayer causes the $C_{60}^{\beta}$ phase to convert to the $C_{60}^{\alpha}$ and bare $C_{60}$ phases. We note that dI/dV spectra of the $C_{60}^{\alpha}$ molecules exhibit pronounced orientational dependence. As seen in Fig. 3(b), the resonances of the 6-6 bond-up molecules are shifted to higher energy with respect to the resonances of the apex-up molecules.

The spatial dependence of the $C_{60}^{\alpha}$ phase molecular electronic states can be seen in energy-resolved spectroscopic *dI/dV* mapping (Fig. 4). The HOMO band of $C_{60}^{\alpha}$ shows highly localized structures, while the LUMO and LUMO+1 *dI/dV* images show long-range filamentary structures that run parallel to the $C_{60}$ 6-6 bonds and are oriented along the Ag[110] direction. We note that the patterns observed in these images primarily reflect a repetition of the topmost portion of the single $C_{60}$ molecular orbitals [17], and are likely not due to a spatial confinement of electrons along this crystallographic direction [5].

Since the 710 K annealed $K_xC_{60}$ film consists of only two phases, a K-free phase and the $C_{60}^{\alpha}$ phase, the stoichiometry of the $C_{60}^{\alpha}$ phase can be determined by assuming that all the deposited K resides in the $C_{60}^{\alpha}$ phase. From the amount of K deposited onto the surface, the packing density of $C_{60}$ molecules in the $C_{60}^{\alpha}$ phase ($0.90 \pm 0.04$ $C_{60}/nm^2$), and the fraction of the surface covered by this phase, the stoichiometry of $C_{60}^{\alpha}$ is found to lie in the $K_xC_{60}$ range of $x = 1 \pm 0.5$. The same analysis for the 640 K annealed film indicates that the $C_{60}^{\beta}$ and $C_{60}^{\alpha}$ phases have nearly identical K concentrations. While we do not observe any changes in the net $C_{60}$ coverage after each anneal, we cannot rule out the possibility that some K is lost during the annealing process, as seen for $K_xC_{60}/Au(111)$[18]. Any K loss would cause our K-doping estimates to be systematically too high for both phases.

From the dI/dV spectra we can estimate the charge state of the $C_{60}$ molecules in the various phases using the simple charge transfer model used in Ref. [19]. This model assumes that electronic spectra rigidly shift upon electron addition [8]. Given the width and degeneracy of the undoped $C_{60}$ LUMO resonance, we assume each added electron shifts the spectrum by approximately -130 meV. We use Gaussian fits of the $C_{60}$ LUMO+1 resonances to track the shifts of the molecular resonances relative to the Fermi energy in the various phases. Changes in the molecular resonances' lineshape and relative energies of the molecular states with K doping limit the accuracy of this model. The spectrum of the undoped $C_{60}$ monolayer indicates very little charge transfer from the Ag(100) substrate (~ 0.1 e-, as estimated in Ref. [17]), evidenced by the lack of spectral density below the Fermi energy. The estimated $C_{60}^{\alpha}$ and $C_{60}^{\beta}$ charge states are $4.1 \pm 0.8$ e- and $2.4 \pm 0.4$ e- per $C_{60}$, respectively. The shifts in spectral density and the implied charge transfer is striking when compared to the charge transfer observed in isolated $C_{60}$ molecules that have been doped with precise numbers of K atoms on bare Ag(001) [19]. The $C_{60}^{\alpha}$ phase molecules appear to have at least one electron more in comparison to an isolated $K_4C_{60}$ complex (i.e. a single $C_{60}$ molecule that is directly attached to four K atoms), despite the fact that the $C_{60}^{\alpha}$ phase has at least two fewer K atoms per $C_{60}$ in comparison [19].

Our most conservative estimates of the $C_{60}^{\alpha}$ monolayer charge state (~ 3 e-) and K-doping concentrations (1.5 K per $C_{60}$) indicate that charge transfer from K alone is too small to account for the difference between the electronic spectra of the $C_{60}^{\alpha}$ phase and undoped $C_{60}$ monolayer phase. A single K atom can donate at most one electron to a $C_{60}^{\alpha}$ molecule, which implies that undoped $C_{60}$ molecules receive at least an additional 1.5 e- of molecular charge in converting to the $C_{60}^{\alpha}$ phase. One possible explanation is that the Ag(001) substrate donates more charge to $C_{60}$ molecules in the $C_{60}^{\alpha}$ monolayer phase than in the undoped monolayer phase. Such charge transfer differences might be the result of differences in substrate bonding between an undoped $C_{60}$ monolayer and a $C_{60}^{\alpha}$ phase monolayer. For example, different bonding characteristics in the undoped $C_{60}$ film lead to observed changes in the $C_{60}$ charge state [9, 16]. Another possibility is that charge differences are exaggerated due to inhomogeneous changes in the spatial distribution of existing occupied electron state density. Since STM probes the local density of states above the molecule, highly inhomogeneous shifts in charge density to/from the $C_{60}$-Ag interface beneath the molecule could potentially lead to incorrect estimates of molecular charge inferred from STS spectra [20]. Such an inhomogeneous state density might explain discrepancies between $C_{60}$ charge transfer values measured using STM and photoemission techniques [6, 8-10].

In summary, we have observed temperature-dependent separation of a $K_xC_{60}$ (x ≈ 1) monolayer into multiple distinct phases which differ in their respective charge states. Higher temperature annealing causes the highest electron-doped phase ($C_{60}^{\alpha}$) to distill out of a lower electron-doped phase ($C_{60}^{\beta}$), leaving behind regions of $C_{60}$ free of K. We find that the $C_{60}^{\alpha}$/Ag(001) monolayer phase exhibits an anomalously high level of occupied molecular state spectral density in comparison to the undoped $C_{60}$/Ag(001) monolayer. The high spectral density could result either from an enhancement of substrate charge transfer or highly inhomogeneous rearrangement of molecular charge caused by K addition.

This work was supported by the Director, Office of Energy Research, Office of Basic Energy Science, Division of Material Sciences and Engineering, U.S. Department of Energy under contract No. DE-AC03-76SF0098.

**Figure Captions**

Figure 1 (a) STM topographs ($V = 2.0$ V, $I = 5$ pA) showing the disordered K-doped $C_{60}$ monolayer after annealing to 415 K  (b) dI/dV spectra of bright and dim $C_{60}$ molecules in the 440 K annealed film (spectra are normalized by the LUMO+1 resonance maximum value). The tip height was stabilized at $V = 2.0$ V, $I = 0.3$ nA for each spectrum.

Figure 2 (a) STM topographs of K-doped $C_{60}$ monolayer annealed to 640 K. The 70x70 Å$^2$ topographs ($V = 2.0$ V, $I = 20$ pA) show the morphology of $C_{60}^{\beta}$ (top right) and $C_{60}^{\alpha}$ (bottom right) phases. The two coexisting phases can be seen in the 2000x2000 Å$^2$ topograph ($V = 2.0$ V, $I = 50$ pA), which is differentiated in the horizontal direction to enhance contrast. A bare Ag(001) region is seen in the center of this image. (b) dI/dV spectra of the $C_{60}^{\alpha}$ and $C_{60}^{\beta}$ phases are shown as the dotted curve ($C_{60}^{\alpha}$ apex atom up molecule, $V = 2.0$ V, $I = 0.5$ nA) and solid curve ($C_{60}^{\beta}$ hexagon up molecule, $V = 2.0$ V, $I = 0.3$ nA). The feature near $V = 1$ V in the $C_{60}^{\beta}$ spectrum is most likely a tip artifact. The curves shown here are spatial averages of spectra acquired over individual molecules and are normalized by the LUMO+1 resonance maximum value.

Figure 3 (a) STM topographs ($V = 2.0$ V, $I = 5$ pA) of the K-doped $C_{60}$ monolayer annealed to 710 K. Close-ups of the $C_{60}^{\alpha}$ and undoped $C_{60}$ phases can be seen in the bottom right and top right topographs, respectively. The 2000x2000 Å$^2$ topograph has been differentiated in the horizontal direction to enhance contrast. (b) dI/dV spectra of the $C_{60}^{\alpha}$ and undoped $C_{60}$ phases are shown as dotted curve ($C_{60}^{\alpha}$ apex atom up molecule, $V = 2.0$ V, $I = 0.3$ nA), short dashed curve ($C_{60}^{\alpha}$ 6-6 bond up molecule, $V = 2.0$ V, $I = 0.5$ nA), long dashed curve (undoped $C_{60}$ dim molecule, $V = 2.0$ V, $I = 0.5$ nA), and solid curve (undoped $C_{60}$ bright molecule, $V = 2.0$ V, $I = 0.5$ nA). The curves shown here are spatial averages of spectra acquired over individual molecules and are normalized by the LUMO+1 resonance maximum value.

Figure 4 STM topograph and energy-resolved $dI/dV$ images of the K-doped $C_{60}^{\alpha}$ phase (120x120 Å$^2$). The topograph was acquired at $V = 1.25$ V, $I = 0.3$ nA. The HOMO, LUMO, and LUMO+1 $dI/dV$ maps were measured at biases of -1.6 V, 0.05 V, and 1.25 V respectively.

Figure 1

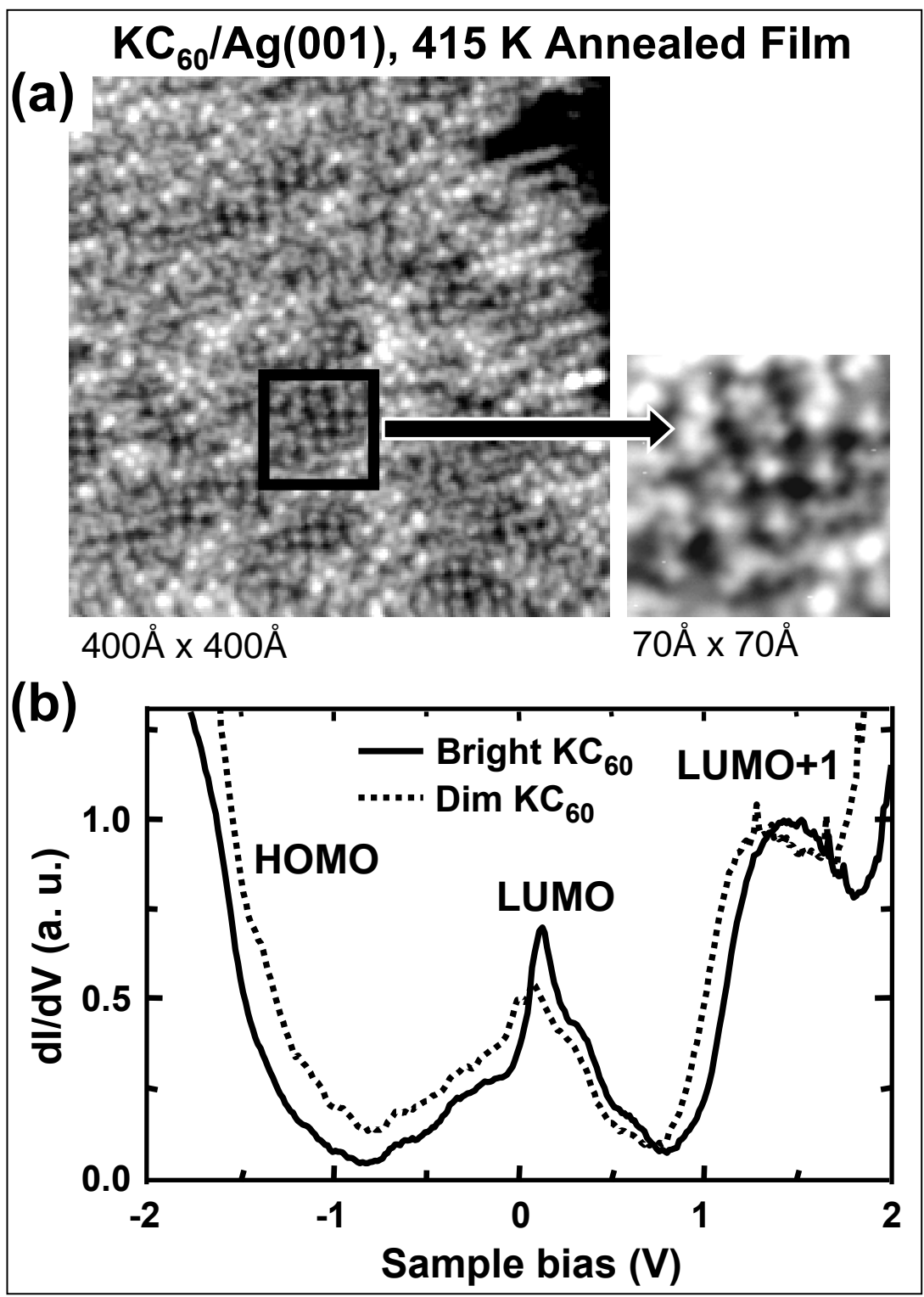

Figure 2

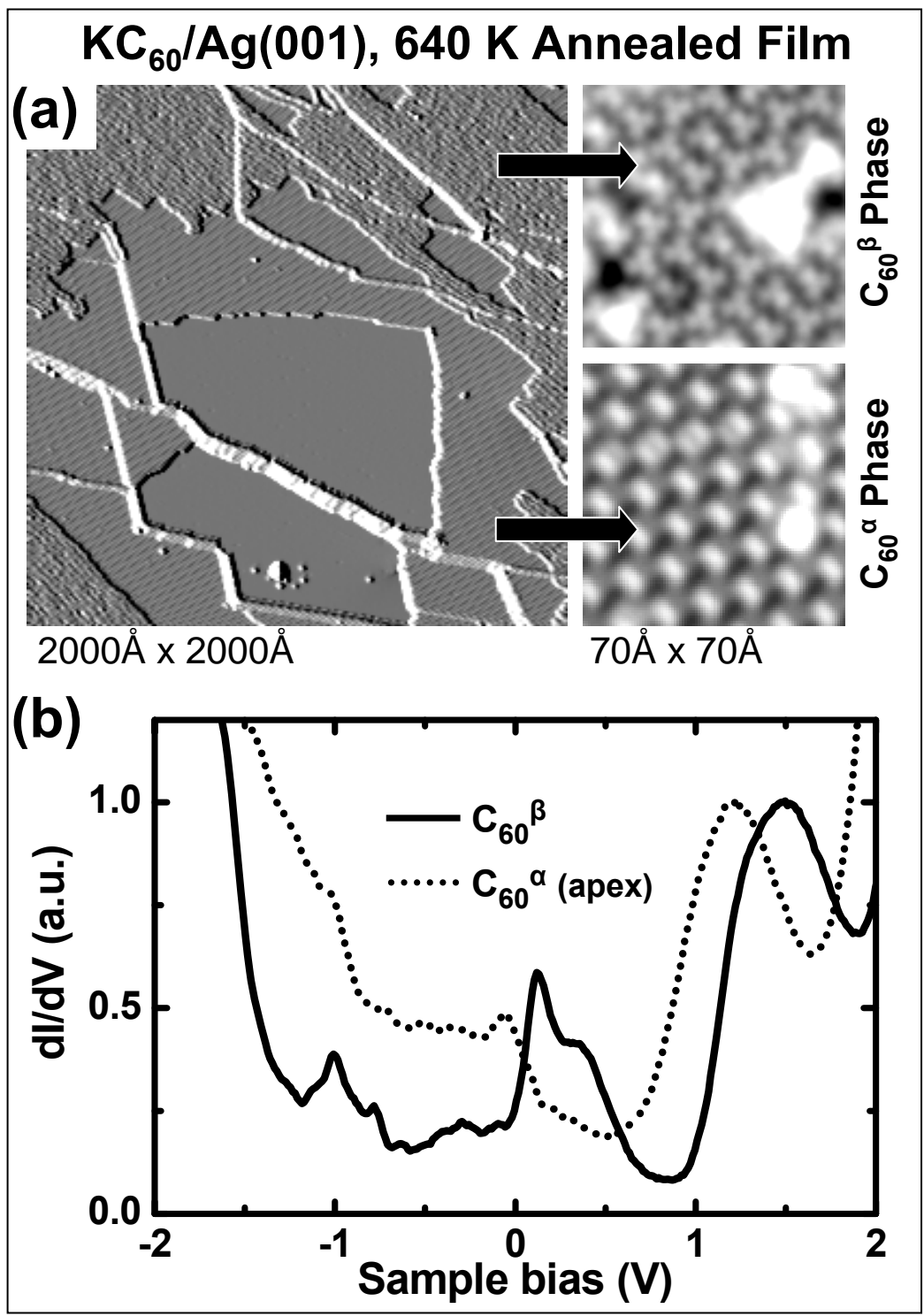

Figure 3

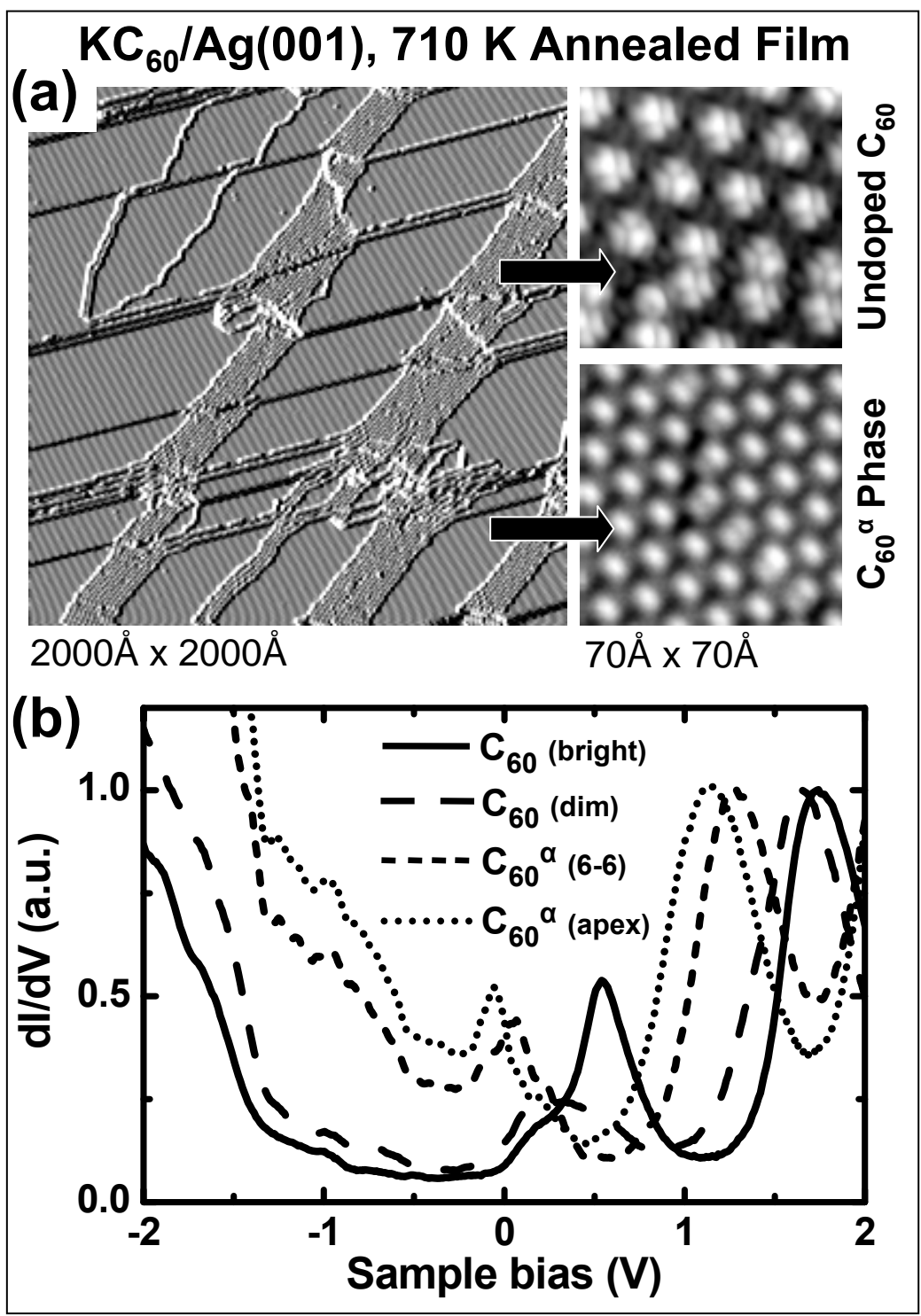

Figure 4

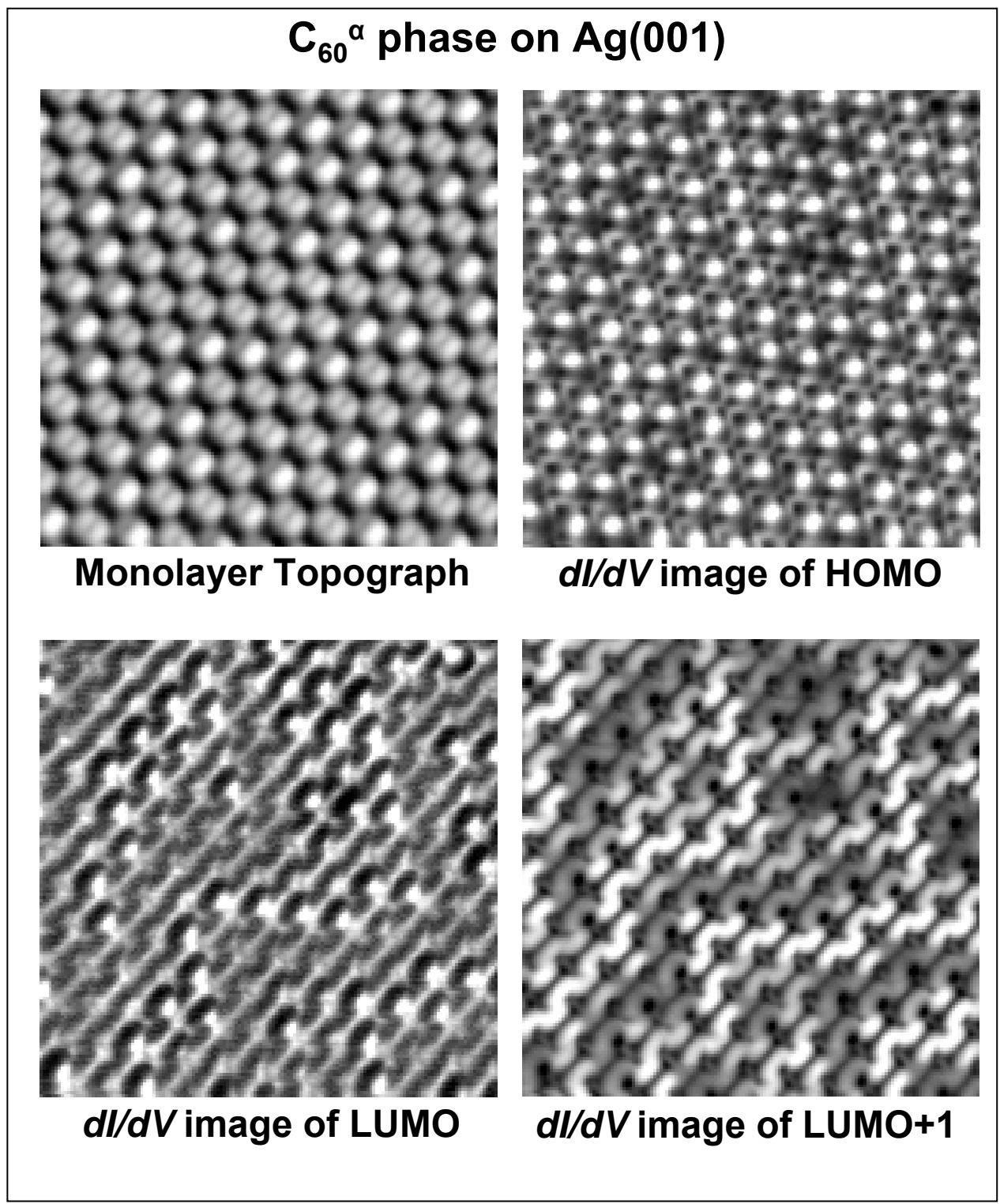